\documentclass[submission,copyright,creativecommons]{eptcs}
\providecommand{\event}{QFM 2012} 

\title{Adding Time to Pushdown Automata \\
{\small (Tutorial)}}
\author{
Parosh Aziz Abdulla
\institute{Department of Information Technology\\ 
Uppsala University\\
Sweden}
\email{parosh@it.uu.se}
\and
Mohamed Faouzi Atig
\institute{Department of Information Technology\\ 
Uppsala University\\
Sweden}
\email{mohamed{\textunderscore}faouzi.atig@it.uu.se}
\and
Jari Stenman
\institute{Department of Information Technology\\ 
Uppsala University\\
Sweden}
\email{jari.stenman@it.uu.se}
}
\def\titlerunning{Adding Time to PDA}
\def\authorrunning{Abdulla et al.}

\usepackage{pgf}
\usepackage{tikz}
\usepackage{clrscode}
\usepackage{amsmath}
\usepackage{amsfonts}
\usepackage{subfig}

\usetikzlibrary{automata}
\usetikzlibrary{petri}
\usetikzlibrary{shadows}
\usetikzlibrary{positioning}
\usetikzlibrary{decorations.pathmorphing}
\usetikzlibrary{decorations.shapes}
\usetikzlibrary{shapes}
\usetikzlibrary{backgrounds}

\usetikzlibrary{fit}
\usetikzlibrary{arrows}
\usetikzlibrary{calc}
\usetikzlibrary{mindmap}
\usetikzlibrary{matrix}
\usepackage{pgffor}
\usepackage{xcolor}

\tikzstyle{west-rsym-node}=[anchor=east,rectangle,rounded corners, fill=blue!30!black!60,text=white,outer sep=0pt,text centered,font=\footnotesize]

\tikzstyle{east-rsym-node}=[anchor=west,rectangle,rounded corners, fill=blue!30!black!60,text=white,outer sep=0pt,text centered,font=\footnotesize]

\tikzstyle{center-rsym-node}=[anchor=center,rectangle,rounded corners, fill=blue!30!black!60,text=white,outer sep=0pt,text centered,font=\footnotesize]

\tikzstyle{dummy-rsym-node}=[anchor=south,rectangle,rounded corners,fill=black!25!blue!25,text=black!25!blue!25,
outer sep=0pt,text centered,font=\footnotesize]

\tikzstyle{empty-rsym-node}=[anchor=east,rectangle,rounded corners,fill=blue!30!black!60,text=blue!30!black!60, 
outer sep=0pt,text centered,font=\footnotesize]

\tikzstyle{stack-sym-node}=[anchor=west,rectangle,rounded corners, fill=black!25!blue!25,outer sep=0pt, inner sep=3pt,
draw=black!70, dashed]

\tikzstyle{stack-node}=[anchor=west,rectangle, fill=black!10!red!15, outer sep=0pt, 
draw=black!70, text=black, font=\footnotesize,text centered]

\tikzstyle{clock-node}=[anchor=west,rectangle, rounded corners, fill=red!60!black!60 ,outer sep=0pt,text=white, font=\footnotesize,text centered]

\tikzstyle{state-node}=[anchor=west,rectangle, rounded corners, fill=red!60!black!60, outer sep=0pt,text=white, font=\footnotesize,text centered]

\tikzstyle{dummy-stack-node}=[anchor=west,rectangle, fill=black!10!red!15,outer sep=0pt, text=black!10!red!15, 
draw=black!70, font=\footnotesize, text centered]

\tikzstyle{label-node}=[font=\footnotesize]

\tikzstyle{tran-edge}=[->,line width=1pt]
\tikzstyle{region-edge}=[color=black!80, dotted, line width=2pt, line cap=round, dash pattern=on .05mm off 0.9mm]
\tikzstyle{region-edge-endpoint}=[fill=none, draw=none, circle, inner sep=0pt, minimum size=4pt]

\newtheorem{theorem}{Theorem}

\newcommand{\tuple}[1]{\left\langle {#1} \right\rangle}

\newcommand{\asym}{a}
\newcommand{\shasym}{a^\bullet}
\newcommand{\bsym}{b}
\newcommand{\shbsym}{b^\bullet}
\newcommand{\csym}{c}

\newcommand{\dsym}{d}

\newcommand{\first}{\vdash}
\newcommand{\shfirst}{\vdash^\bullet}
\newcommand{\xclock}{x}
\newcommand{\shxclock}{x^\bullet}
\newcommand{\yclock}{y}
\newcommand{\shyclock}{y^\bullet}

\newcommand{\state}{s}
\newcommand{\conf}{c}

\newcommand{\mset}[1]{\lbrace #1 \rbrace}
\newcommand{\nnreals}{\mathbb{R}^{\geq 0}}

\newcommand{\nop}{\mathbf{nop}}
\newcommand{\op}{\mathbf{op}}

\newcommand{\push}[1]{\mathbf{push}({#1})}
\newcommand{\pop}[1]{\mathbf{pop}({#1})}

\newcommand{\pusht}[2]{\mathbf{push}({#1}, {#2})}
\newcommand{\popt}[2]{\mathbf{pop}({#1}, {#2})}

\newcommand{\test}[2]{{#1} \in {#2}\,?}
\newcommand{\reset}[2]{{#1} \gets {#2}}

\newcommand{\ppushop}[1]{\push{#1}}
\newcommand{\ppopop}[1]{\pop{#1}}

\begin{document}
\maketitle

\begin{abstract}
In this tutorial, we illustrate through examples how
we can combine two classical models, namely those 
of {\it pushdown automata} ({\sc Pda}) and
{\it timed automata}, in order to obtain 
{\it timed pushdown automata} ({\sc Tpda})
\cite{abdulla2012minimal,abdulla2012dense}.
Furthermore, we describe how the reachability problem for {\sc Tpda}s can be 
reduced to the reachability problem for {\sc Pda}s.

\end{abstract}


\section{Introduction}

In this tutorial, we describe a timed extension of the widely used model of
 Pushdown Automata ({\sc Pda}) \cite{abdulla2012minimal,abdulla2012dense}.
A {\sc Pda} computes by moving between states according to some given transition rules. Additionally,
a {\sc Pda} may utilize a stack to store information. This information is encoded in \emph{stack symbols}, 
and the {\sc Pda} may add a symbol (\emph{push}) to or remove a symbol (\emph{pop}) from the stack. The defining feature of a stack is that it
has ordering on its elements, traditionally from \emph{top} to \emph{bottom}; the {\sc Pda} can only access the topmost element.

An interesting question is what happens to this model when we extend it with quantitative properties. 
Will basic problems, such as state reachability, still be decidable? In particular, we are interested in
extending the model with continuous time in a similar manner in which Timed Automata \cite{alur1994theory} extend Finite Automata.
Thus, we consider Timed Pushdown automata
{\sc Tpda}.
A {\sc Tpda} is a {\sc Pda} that is augmented with a finite number of \emph{clocks}.
It operates in the following manner:

\begin{itemize}
\item at any point in the computation, time may elapse by some real number, increasing the values of all clocks
\item the values of clocks constrain the actions of the automaton 
\end{itemize}

In addition to the set of clocks, we also store the age of each stack symbol. 
We can view this as an additional clock. Accordingly, the ages of stack symbols increase whenever time elapses.
Furthermore, possible actions of the automaton may be restricted by the age of topmost stack symbol.

The {\sc Tpda}  model thus subsumes both the model
of pushdown automata and timed automata.
More precisely, we obtain the former
if we prevent the
{\sc Tpda} from using the timed information (all the
timing constraints are trivially valid); and
obtain the latter if we prevent the
{\sc Tpda} from using the stack 
(no symbols are pushed to or popped from the stack).
Notice that a {\sc Tpda} induces a system that is infinite 
in two dimensions, namely
it gives rise to a stack containing an unbounded number
of symbols each of which is equipped with a real-valued
clock.

\paragraph{Outline}

In the next section, we present an overview of Pushdown Automata. 
In Section \ref{sec:tpda}, we describe the timed extension of {\sc Pda}
 and show some examples of computations.
In Section \ref{sec:regions}, we recall and extend the notion of regions, and show how 
we can use them to define a symbolic encoding of {\sc Tpda} configurations.
Finally, in Section \ref{sec:translation} we describe how to construct a {\sc Pda} which simulates a given {\sc Tpda}.
The section ends with a detailed example of how the aforementioned {\sc Tpda} computation is simulated.


\section{{\sc Pda}} \label{sec:pda}

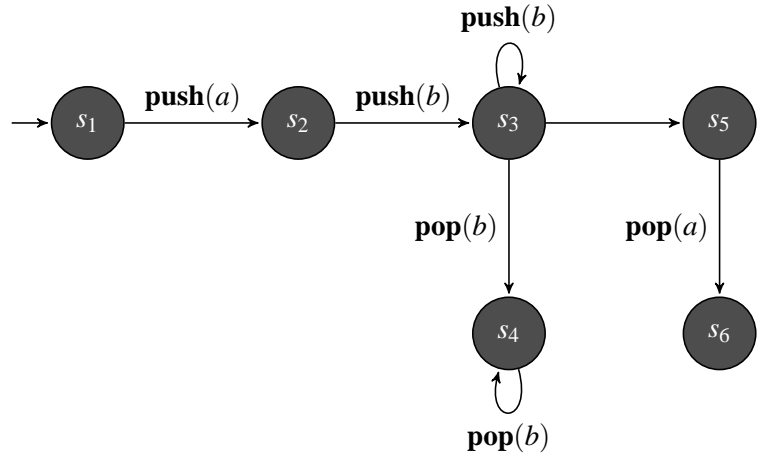
\begin{figure}
\centering
\begin{tikzpicture}[->,>=stealth',shorten >=1pt,auto,node distance=2.8cm,semithick, initial text=]

\tikzstyle{every state}=[fill=black!70,draw=black,text=white]

\node[state, initial] (s1) {$s_1$};
\node[state] (s2) [right of=s1] {$s_2$};
\node[state] (s3) [right of=s2] {$s_3$};
\node[state] (s4) [right of=s3] {$s_5$};
\node[state] (s5) [below of=s3] {$s_4$};
\node[state] (s6) [below of=s4] {$s_6$};

\path (s1) edge            node {$\push{a}$} (s2)
      (s2) edge            node {$\push{b}$} (s3)
      (s3) edge [loop above] node {$\push{b}$} (s3)
           edge            node {} (s4)
      (s3) edge node [left] {$\pop{b}$} (s5)
      (s4) edge node [left] {$\pop{a}$} (s6)
      (s5) edge [loop below] node {$\pop{b}$} (s5);

\end{tikzpicture}
\caption{A simple PDA}
\label{fig:simple_pda}
\end{figure}

In this section, we informally describe the model of Pushdown Automata. 
A Pushdown Automaton (PDA) is a tuple $(S, s_{\it init}, \Gamma, \Delta)$ consisting of a finite set
of \emph{states} $S$, an initial state $s_{\it init}$,  a finite \emph{stack alphabet} $\Gamma$,
and a finite set of \emph{transition rules} $\Delta$. 
During the operation of a {\sc Pda}, it may store information in a stack.
It may add information, which is referred to as \emph{pushing}, or it may remove information, which is called \emph{popping}.
The stack is a last-in, first-out queue, and access is restricted to the first element.
The stack alphabet contains all possible symbols that may be stored in the stack,
and the set of transition rules describe the manner
in which the automaton is allowed to move between states.
Each transition rule is of the form $(s, \op, t)$. The rule contains a source state $s$, a target state $t$ and a stack operation $\op$.
The stack operation is either $\push{a}$, $\pop{a}$ or $\nop$ (here, $a$ is an arbitrary symbol from the stack alphabet).
A transition rule describes that the automaton may move from $s$ to $t$ while performing the stack operation $\op$. 
The operation $\push{a}$ pushes $a$ onto the stack, and $\pop{a}$ pops it. The operation $\nop$ is 
an ``empty'' operation which can be used to change state without modifying the stack.
Figure~\ref{fig:simple_pda} shows a PDA with the state set $\mset{s_1,s_2,s_3,s_4,s_5,s_6}$ and stack alphabet $\mset{a,b}$.
The initial state of the automaton is $s_1$.
The transition rules are drawn as arrows between states, labeled with the stack operation (missing labels mean $\nop$).



\begin{figure}[h]
\begin{tikzpicture}

\node [name=ref] {};

\node[state-node,name=state-node]
at (ref.west)
{
$
\begin{array}{c}
\state_1
\end{array}
$
};

\node[dummy-stack-node,name=stack-node] 
at ($(state-node.east)+(1mm,0mm)$)
{
$
\begin{array}{c}
a
\end{array}
$
};

\node[anchor=north,name=conf-node,inner sep=0mm, outer sep=0mm] at ($(state-node.south)+(-1mm,-2mm)$) {$\conf_0$};

\begin{pgfonlayer}{background}
\node[fill=yellow!20,draw,rounded corners,name=conf-area,fit= (state-node)   (stack-node) (conf-node)]{};

\end{pgfonlayer}

\draw[tran-edge] ($(conf-area.east)+(2.5mm,0mm)$) -- node[label-node,above=2pt] {$\ppushop{a}$} ($(conf-area.east)+(20mm,0mm)$);

\node[state-node,name=state-node]
at ($(stack-node.east)+(27mm,0mm)$) 
{
$
\begin{array}{c}
\state_2
\end{array}
$
};

\node[stack-node,name=stack-node] 
at ($(state-node.east)+(1mm,0mm)$)
{
$
\begin{array}{c}
a
\end{array}
$
};

\node[anchor=north,name=conf-node,inner sep=0mm, outer sep=0mm] at ($(state-node.south)+(-1mm,-2mm)$) {$\conf_1$};

\begin{pgfonlayer}{background}
\node[fill=yellow!20,draw,rounded corners,name=conf-area,fit= (state-node)   (stack-node) (conf-node)]{};

\end{pgfonlayer}

\draw[tran-edge] ($(conf-area.east)+(2.5mm,0mm)$) -- node[label-node,above=2pt] {$\ppushop{b}$} ($(conf-area.east)+(20mm,0mm)$);

\node[state-node,name=state-node]
at ($(stack-node.east)+(27mm,0mm)$) 
{
$
\begin{array}{c}
\state_3
\end{array}
$
};

\node[stack-node,name=stack-node] 
at ($(state-node.east)+(1mm,0mm)$)
{
$
\begin{array}{c}
b\\a
\end{array}
$
};

\node[anchor=north,name=conf-node,inner sep=0mm, outer sep=0mm] at ($(state-node.south)+(-1mm,-2mm)$) {$\conf_2$};

\begin{pgfonlayer}{background}
\node[fill=yellow!20,draw,rounded corners,name=conf-area,fit= (state-node)   (stack-node) (conf-node)]{};

\end{pgfonlayer}

\draw[tran-edge] ($(conf-area.east)+(2.5mm,0mm)$) -- node[label-node,above=2pt] {$\ppopop{b}$} ($(conf-area.east)+(20mm,0mm)$);

\node[state-node,name=state-node]
at ($(stack-node.east)+(27mm,0mm)$) 
{
$
\begin{array}{c}
\state_4
\end{array}
$
};

\node[stack-node,name=stack-node] 
at ($(state-node.east)+(1mm,0mm)$)
{
$
\begin{array}{c}
a
\end{array}
$
};

\node[anchor=north,name=conf-node,inner sep=0mm, outer sep=0mm] at ($(state-node.south)+(-1mm,-2mm)$) {$\conf_3$};

\begin{pgfonlayer}{background}
\node[fill=yellow!20,draw,rounded corners,name=conf-area,fit= (state-node)   (stack-node) (conf-node)]{};

\end{pgfonlayer}

\end{tikzpicture}
\caption{Computation of a PDA}
\label{fig:example_pda_computation}
\end{figure}
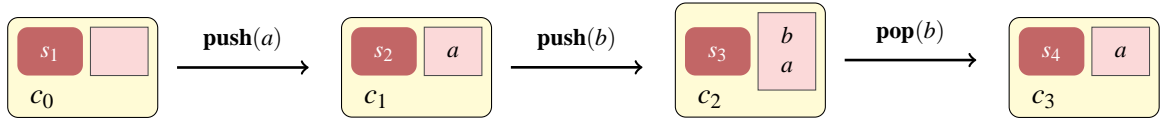

At any point during a computation, the PDA is in a certain \emph{configuration}, defined by the current state and the current stack content. 
Figure \ref{fig:example_pda_computation} shows the configurations that appear along a computation in which
the automaton starts from its initial configuration (the state is $s_1$ and the stack is empty), 
moves to $s_2$ while pushing $a$, 
then moves to $s_3$ while pushing $b$, and finally pops $b$ and moves to $s_4$.





\paragraph{Reachability} Given a pushdown automaton, 
the reachability problem is the problem of deciding whether the automaton can reach a particular state $s$.
In other words, 
we ask whether there is a computation of the automaton 
(starting from the initial configuration) that visits a configuration
where the state is $s$, regardless of the content of the stack.
It turns out that for the automaton in Figure~\ref{fig:simple_pda}, the state $s_4$ is reachable but the state $s_6$ is not.
This is because in order to move from $s_5$ to $s_6$, the automaton has to pop $a$. However, the topmost symbol when
the automaton is in state $s_5$ will always be $b$.
For PDA, reachability is decidable in polynomial time \cite{BEM97}.


\section{Timed Pushdown Automata}\label{sec:tpda}



The classical model of Timed Automata extends finite state automata with a finite set of real-valued \emph{clocks}.
We extend {\sc Pda} in a similar way, in the sense that a Timed Pushdown Automaton ({\sc Tpda}) consists of a finite set of states $S$,
an initial state $s_{\it init}$,
a finite stack alphabet $\Gamma$, a finite set of transition rules $\Delta$, and a finite set of clocks $X$.
The transition rules are also extended in the sense that they can read and write the values of clocks.
More specifically, a transition rule $(s, \op, t)$ refers not only to stack operations.
Instead, $\op$ can also be one of the \emph{clock operations} $\test{x}{I}$ and $\reset{x}{I}$.
The operation $\test{x}{I}$ checks whether the value of the clock $x$ is in the interval $I$.
For example, the transition rule $(s, \test{x}{[1:3]}, t)$ can only be performed when the value of $x$ is between 1 and 3.
The operation $\reset{x}{I}$ \emph{nondeterministically} resets the value of the clock $x$ to some value in the interval $I$.
Additionally, each stack symbol is equipped with a value representing its \emph{age}.
We modify the stack operations to use these values: $\pusht{\asym}{I}$ pushes $\asym$ and nondeterministically 
sets its initial age to some value in the interval $I$, while $\popt{\asym}{I}$ may only pop the topmost stack symbol
if it is equal to $\asym$ and its age is in the given interval $I$.

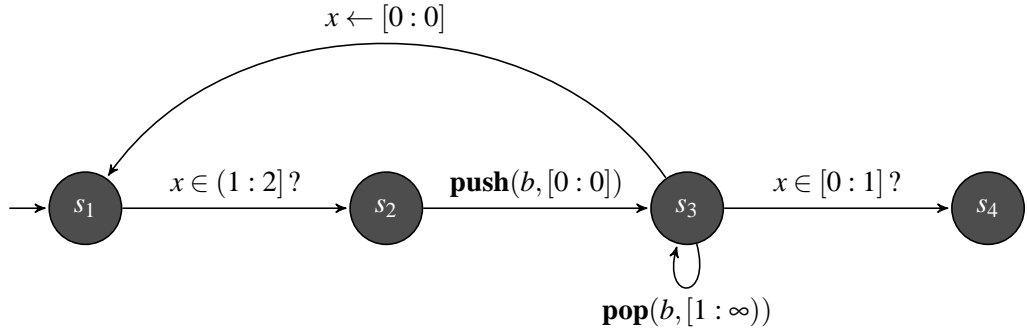
\begin{figure}
\centering
\begin{tikzpicture}[->,>=stealth',shorten >=1pt,auto,node distance=4cm,semithick, initial text=]

\tikzstyle{every state}=[fill=black!70,draw=black,text=white]

\node[state, initial] (s1) {$s_1$};
\node[state] (s2) [right of=s1] {$s_2$};
\node[state] (s3) [right of=s2] {$s_3$};
\node[state] (s4) [right of=s3] {$s_4$};

\path (s1) edge            node {$\test{x}{(1 : 2]}$} (s2)
      (s2) edge            node {$\pusht{b}{[0 : 0]}$} (s3)
      (s3) edge [loop below] node {$\popt{b}{[1 : \infty)}$} (s3)
      (s3) edge            node {$\test{x}{[0 : 1]}$} (s4)
      (s3) edge [bend right=55] node [above] {$\reset{x}{[0 : 0]}$} (s1);

\end{tikzpicture}
\caption{A simple TPDA}
\label{fig:simple_tpda}
\end{figure}
\begin{figure}
\begin{tikzpicture}

\node [name=ref] {};

\node[state-node,name=state-node]
at (ref.west)
{
$
\begin{array}{c}
\state_1
\end{array}
$
};

\node[clock-node,name=clock-node]
at ($(state-node.east)+(1mm,0mm)$)
{
$
\begin{array}{c}
\xclock\gets 2.1\\
\yclock\gets 5.9\\
\end{array}
$
};
\node[stack-node,name=stack-node] 
at ($(clock-node.east)+(1mm,0mm)$)
{
$
\begin{array}{c}
\tuple{\csym,6.1}\\             
\tuple{\bsym,8.2}\\             
\tuple{\asym,9.9}
\end{array}
$
};

\node[anchor=north,name=conf-node,inner sep=0mm, outer sep=0mm] at ($(state-node.south)+(-1mm,-2mm)$) {$\conf_{13}$};

\begin{pgfonlayer}{background}
\node[fill=yellow!20,draw,rounded corners,name=conf-area,fit= (state-node) (clock-node)  (stack-node) (conf-node)]{};

\end{pgfonlayer}

\draw[tran-edge] ($(conf-area.east)+(2.5mm,0mm)$) -- node[label-node,above=2pt] 
{$\popt{c}{(6:\infty)}$} ($(conf-area.east)+(20mm,0mm)$);

\node[state-node,name=state-node]
at ($(conf-area.east)+(25mm,0mm)$) 
{
$
\begin{array}{c}
\state_1
\end{array}
$
};

\node[clock-node,name=clock-node]
at ($(state-node.east)+(1mm,0mm)$)
{
$
\begin{array}{c}
\xclock\gets 2.1\\
\yclock\gets 5.9\\
\end{array}
$
};
\node[stack-node,name=stack-node] 
at ($(clock-node.east)+(1mm,0mm)$)
{
$
\begin{array}{c}
\tuple{\bsym,8.2}\\             
\tuple{\asym,9.9}
\end{array}
$
};

\node[anchor=north,name=conf-node,inner sep=0mm, outer sep=0mm] at ($(state-node.south)+(-1mm,-2mm)$) {$\conf_{14}$};

\begin{pgfonlayer}{background}
\node[fill=yellow!20,draw,rounded corners,name=conf-area,fit= (state-node) (clock-node)  (stack-node) (conf-node)]{};

\end{pgfonlayer}

\draw[tran-edge] ($(conf-area.east)+(2.5mm,0mm)$) -- node[label-node,above=2pt] 
{$\xclock\gets (2:3)$} ($(conf-area.east)+(20mm,0mm)$);

\node[state-node,name=state-node]
at ($(ref.west) + (0mm,-22mm)$)
{
$
\begin{array}{c}
\state_3
\end{array}
$
};

\node[clock-node,name=clock-node]
at ($(state-node.east)+(1mm,0mm)$)
{
$
\begin{array}{c}
\xclock\gets 2.2\\
\yclock\gets 5.9\\
\end{array}
$
};
\node[stack-node,name=stack-node] 
at ($(clock-node.east)+(1mm,0mm)$)
{
$
\begin{array}{c}
\tuple{\bsym,8.2}\\             
\tuple{\asym,9.9}
\end{array}
$
};

\node[anchor=north,name=conf-node,inner sep=0mm, outer sep=0mm] at ($(state-node.south)+(-1mm,-2mm)$) {$\conf_{15}$};

\begin{pgfonlayer}{background}
\node[fill=yellow!20,draw,rounded corners,name=conf-area,fit= (state-node) (clock-node)  (stack-node) (conf-node)]{};

\end{pgfonlayer}

\draw[tran-edge] ($(conf-area.east)+(2.5mm,0mm)$) -- node[label-node,above=2pt] 
{$\yclock\gets (0:1)$} ($(conf-area.east)+(20mm,0mm)$);

\node[state-node,name=state-node]
at ($(conf-area.east)+(25mm,0mm)$) 
{
$
\begin{array}{c}
\state_4
\end{array}
$
};

\node[clock-node,name=clock-node]
at ($(state-node.east)+(1mm,0mm)$)
{
$
\begin{array}{c}
\xclock\gets 2.2\\
\yclock\gets 0.4\\
\end{array}
$
};
\node[stack-node,name=stack-node] 
at ($(clock-node.east)+(1mm,0mm)$)
{
$
\begin{array}{c}
\tuple{\bsym,8.2}\\             
\tuple{\asym,9.9}
\end{array}
$
};

\node[anchor=north,name=conf-node,inner sep=0mm, outer sep=0mm] at ($(state-node.south)+(-1mm,-2mm)$) {$\conf_{16}$};

\begin{pgfonlayer}{background}
\node[fill=yellow!20,draw,rounded corners,name=conf-area,fit= (state-node) (clock-node)  (stack-node) (conf-node)]{};

\end{pgfonlayer}

\draw[tran-edge] ($(conf-area.east)+(2.5mm,0mm)$) -- node[label-node,above=2pt] 
{$\pusht{\dsym}{[1:5]}$} ($(conf-area.east)+(20mm,0mm)$);

\node[state-node,name=state-node]
at ($(ref.west) + (0mm,-48mm)$)
{
$
\begin{array}{c}
\state_2
\end{array}
$
};

\node[clock-node,name=clock-node]
at ($(state-node.east)+(1mm,0mm)$)
{
$
\begin{array}{c}
\xclock\gets 2.2\\
\yclock\gets 0.4\\
\end{array}
$
};
\node[stack-node,name=stack-node] 
at ($(clock-node.east)+(1mm,0mm)$)
{
$
\begin{array}{c}
\tuple{\dsym,2.3}\\             
\tuple{\bsym,8.2}\\             
\tuple{\asym,9.9}
\end{array}
$
};

\node[anchor=north,name=conf-node,inner sep=0mm, outer sep=0mm] at ($(state-node.south)+(-1mm,-2mm)$) {$\conf_{17}$};

\begin{pgfonlayer}{background}
\node[fill=yellow!20,draw,rounded corners,name=conf-area,fit= (state-node) (clock-node)  (stack-node) (conf-node)]{};

\end{pgfonlayer}

\draw[tran-edge] ($(conf-area.east)+(2.5mm,0mm)$) -- node[label-node,above=2pt] 
{$\xclock\gets[0:2]$} ($(conf-area.east)+(20mm,0mm)$);

\node[state-node,name=state-node]
at ($(conf-area.east)+(25mm,0mm)$) 
{
$
\begin{array}{c}
\state_4
\end{array}
$
};

\node[clock-node,name=clock-node]
at ($(state-node.east)+(1mm,0mm)$)
{
$
\begin{array}{c}
\xclock\gets 0.3\\
\yclock\gets 0.4\\
\end{array}
$
};
\node[stack-node,name=stack-node] 
at ($(clock-node.east)+(1mm,0mm)$)
{
$
\begin{array}{c}
\tuple{\dsym,2.3}\\             
\tuple{\bsym,8.2}\\             
\tuple{\asym,9.9}
\end{array}
$
};

\node[anchor=north,name=conf-node,inner sep=0mm, outer sep=0mm] at ($(state-node.south)+(-1mm,-2mm)$) {$\conf_{18}$};

\begin{pgfonlayer}{background}
\node[fill=yellow!20,draw,rounded corners,name=conf-area,fit= (state-node) (clock-node)  (stack-node) (conf-node)]{};

\end{pgfonlayer}

\draw[tran-edge] ($(conf-area.east)+(2.5mm,0mm)$) -- node[label-node,above=2pt] 
{${\it Time}=1.75$} ($(conf-area.east)+(20mm,0mm)$);

\node[state-node,name=state-node]
at ($(ref.west) + (0mm,-74mm)$)
{
$
\begin{array}{c}
\state_1
\end{array}
$
};

\node[clock-node,name=clock-node]
at ($(state-node.east)+(1mm,0mm)$)
{
$
\begin{array}{c}
\xclock\gets 2.05\\
\yclock\gets 2.15\\
\end{array}
$
};
\node[stack-node,name=stack-node] 
at ($(clock-node.east)+(1mm,0mm)$)
{
$
\begin{array}{c}
\tuple{\dsym,4.05}\\             
\tuple{\bsym,9.95}\\             
\tuple{\asym,11.85}
\end{array}
$
};

\node[anchor=north,name=conf-node,inner sep=0mm, outer sep=0mm] at ($(state-node.south)+(-1mm,-2mm)$) {$\conf_{19}$};

\begin{pgfonlayer}{background}
\node[fill=yellow!20,draw,rounded corners,name=conf-area,fit= (state-node) (clock-node)  (stack-node) (conf-node)]{};

\end{pgfonlayer}

\draw[tran-edge] ($(conf-area.east)+(2.5mm,0mm)$) -- node[label-node,above=2pt] 
{$\popt{\dsym}{[4:5)}$} ($(conf-area.east)+(20mm,0mm)$);

\node[state-node,name=state-node]
at ($(conf-area.east)+(25mm,0mm)$) 
{
$
\begin{array}{c}
\state_1
\end{array}
$
};

\node[clock-node,name=clock-node]
at ($(state-node.east)+(1mm,0mm)$)
{
$
\begin{array}{c}
\xclock\gets 2.05\\
\yclock\gets 2.15\\
\end{array}
$
};
\node[stack-node,name=stack-node] 
at ($(clock-node.east)+(1mm,0mm)$)
{
$
\begin{array}{c}
\tuple{\bsym,9.95}\\
\tuple{\asym,11.85}
\end{array}
$
};

\node[anchor=north,name=conf-node,inner sep=0mm, outer sep=0mm] at ($(state-node.south)+(-1mm,-2mm)$) {$\conf_{20}$};

\begin{pgfonlayer}{background}
\node[fill=yellow!20,draw,rounded corners,name=conf-area,fit= (state-node) (clock-node)  (stack-node) (conf-node)]{};

\end{pgfonlayer}

\draw[tran-edge] ($(conf-area.east)+(2.5mm,0mm)$) -- node[label-node,above=2pt] 
{$\xclock\gets(3:4)$} ($(conf-area.east)+(20mm,0mm)$);

\node[state-node,name=state-node]
at ($(ref.west) + (0mm,-96mm)$)
{
$
\begin{array}{c}
\state_1
\end{array}
$
};

\node[clock-node,name=clock-node]
at ($(state-node.east)+(1mm,0mm)$)
{
$
\begin{array}{c}
\xclock\gets 3.05\\
\yclock\gets 2.15\\
\end{array}
$
};
\node[stack-node,name=stack-node] 
at ($(clock-node.east)+(1mm,0mm)$)
{
$
\begin{array}{c}
\tuple{\bsym,9.95}\\
\tuple{\asym,11.85}
\end{array}
$
};

\node[anchor=north,name=conf-node,inner sep=0mm, outer sep=0mm] at ($(state-node.south)+(-1mm,-2mm)$) {$\conf_{21}$};

\begin{pgfonlayer}{background}
\node[fill=yellow!20,draw,rounded corners,name=conf-area,fit= (state-node) (clock-node)  (stack-node) (conf-node)]{};

\end{pgfonlayer}

\draw[tran-edge] ($(conf-area.east)+(2.5mm,0mm)$) -- node[label-node,above=2pt] 
{$\test{y}{(1:\infty)}$} ($(conf-area.east)+(20mm,0mm)$);

\node[state-node,name=state-node]
at ($(conf-area.east)+(25mm,0mm)$) 
{
$
\begin{array}{c}
\state_5
\end{array}
$
};

\node[clock-node,name=clock-node]
at ($(state-node.east)+(1mm,0mm)$)
{
$
\begin{array}{c}
\xclock\gets 3.05\\
\yclock\gets 2.15\\
\end{array}
$
};
\node[stack-node,name=stack-node] 
at ($(clock-node.east)+(1mm,0mm)$)
{
$
\begin{array}{c}
\tuple{\bsym,9.95}\\
\tuple{\asym,11.85}
\end{array}
$
};

\node[anchor=north,name=conf-node,inner sep=0mm, outer sep=0mm] at ($(state-node.south)+(-1mm,-2mm)$) {$\conf_{22}$};

\begin{pgfonlayer}{background}
\node[fill=yellow!20,draw,rounded corners,name=conf-area,fit= (state-node) (clock-node)  (stack-node) (conf-node)]{};

\end{pgfonlayer}

\end{tikzpicture}
\caption{A computation of a TPDA (continued)}
\label{fig:tpda_example_2}
\end{figure}
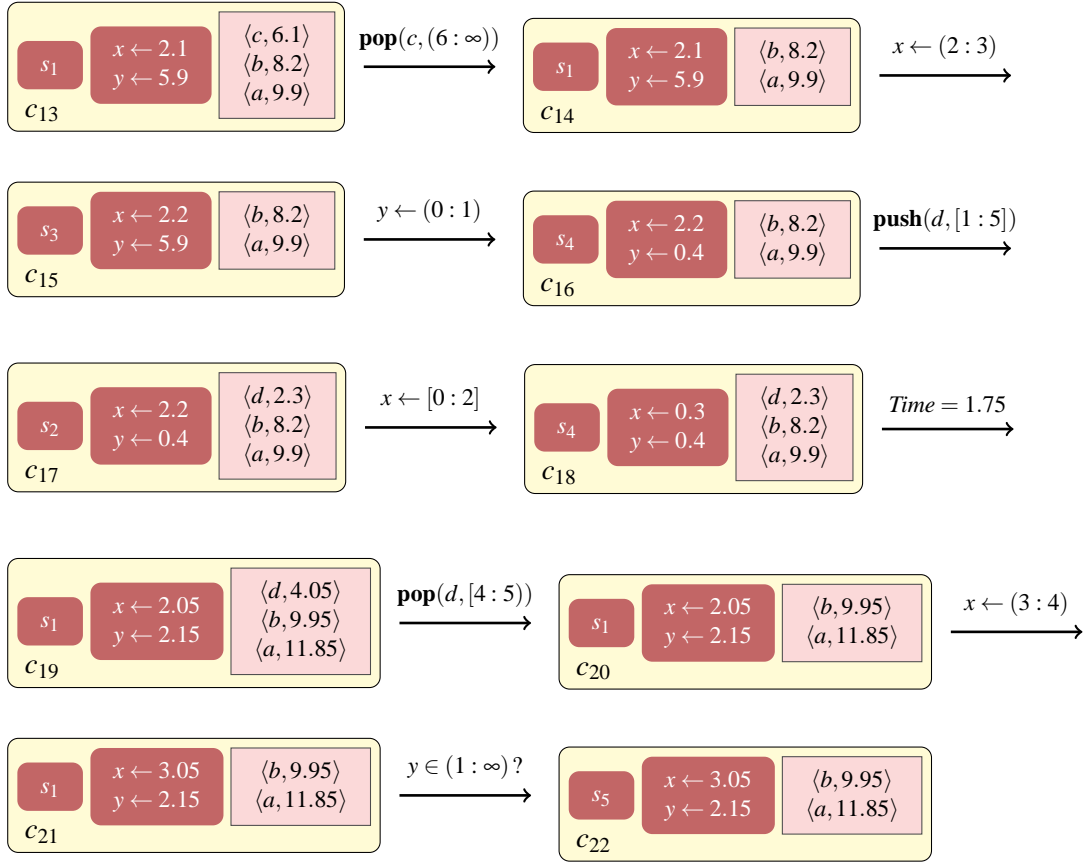

As with {\sc Pda}, the semantics of {\sc Tpda} are given by a transition system over configurations. 
The configurations of a {\sc Tpda} need to contain additional information, namely the values of all clocks and the ages of all stack symbols.
The values of all clocks are given by a \emph{clock valuation}; a mapping $X \mapsto \nnreals$ 
(where $\nnreals$ stands for the non-negative real numbers). 
To capture the ages of clocks symbols, we store tuples in the stack.
Each tuple consists of (i) a stack symbol from the stack alphabet $\Gamma$ and (ii) its corresponding age.
Figure \ref{fig:tpda_example_1} and Figure \ref{fig:tpda_example_2} show an example computation of a {\sc Tpda}
(note that this computation is not related to the automaton in Figure \ref{fig:simple_tpda}).
For example, in the configuration $c_0$ in Figure \ref{fig:tpda_example_1}, the automaton is in the state $s_1$ with an empty stack,
and the values of the two clocks $x$ and $y$ are 0.
In the configuration $c_3$ in the same figure, the stack consists of a symbol $a$ which has age $2.4$.

There are two different types of transitions between configurations of a {\sc Tpda}; \emph{discrete} and \emph{timed}.
Discrete transitions are direct applications of the transition rules in $\Delta$. 
Timed transitions simulate the passage of time. At any point in the computation, the automaton may take a timed transition, which means
that all clock values and ages of stack symbols are increased by a positive real number.
Figures \ref{fig:tpda_example_1} and \ref{fig:tpda_example_2} show a computation of a {\sc Tpda} with clocks $X = \mset{x,y}$ and
stack alphabet $\Gamma = \mset{a,b,c,d}$. We will describe the effect of each type of transition with an example from 
these figures.

Between $c_2$ and $c_3$, the {\sc Tpda} moves from $s_2$ to $s_3$ and pushes the symbol $a$ onto an empty stack, setting its
initial age to $2.4$, a value which is in the allowed interval $[1:3)$. Recall that the initial age is nondeterministically 
chosen from the given interval; in the push between $c_6$ and $c_7$ the same interval is given, but the chosen value happens to be 
$2.9$ instead. The operation $\reset{x}{I}$ chooses and assigns a value nondeterministically. 
From $c_7$, the automaton resets the value of $x$.
Its value, which was previously $6.1$, is set to some value in the interval $[2:3]$, in this case $2.1$.
Assume that $\Delta$ contains a transition rule $(s_1, \test{y}{(1:\infty)}, s_5)$.
In $c_{21}$, the {\sc Tpda} tests if the value of $y$ is strictly greater than $1$. It is, so the transition rule is applied, 
and the state changes to $s_5$, as shown in configuration $c_{22}$.
The above transitions are all examples of discrete transitions, i.e. transitions that are induced by transition rules in $\Delta$.
Figure \ref{fig:tpda_example_1} and Figure \ref{fig:tpda_example_2} also
contain a number of timed transitions.
For example, the transition between $c_8$ and $c_9$ represents the passage of $0.9$ time units.
In $c_9$, the values of $x$ and $y$ and the ages of $a$ and $b$ have all been increased by $0.9$.


\input{tpda_example_1}


\paragraph{Reachability}

In a similar manner to the reachability problem for {\sc Pda}, 
the reachability problem for {\sc Tpda} is the problem of deciding whether a particular state is reachable from the 
initial configuration or not. 
In other words, we ask whether it is possible to reach a configuration $c$ 
such that the state of $c$ is the given target state.

Notice that in the definition of the reachability problem, we do not place any restrictions on
the stack contents or on the values of the clocks.
However, the reachability of a state in a {\sc Tpda} may, in general, depend on the clock values
and the ages of the stack symbols.
For example, the state $s_4$ in Figure \ref{fig:simple_tpda} is not reachable because of timing limitations.

Since the set of configurations in a {\sc Tpda} is infinite,
we can not solve the reachability problem by iteratively computing the successors of the initial configuration
until a fixed point is reached. Furthermore, we cannot use the classical techniques that solve the reachability problem for 
{\sc Pda} \cite{BEM97} since those constructions rely on the stack alphabet being finite.
Therefore, we will now describe a symbolic representation of clock valuations and 
ages of stack symbols. We will use this representation to construct a \emph{symbolic} 
{\sc Pda} that simulates the behavior of the given {\sc Tpda}.


\section{Regions}
\label{sec:regions}

In this section, we describe a symbolic region encoding to represent the infinitely many 
clock valuations of a {\sc Tpda} in a finite way. 
In the following section, we show how to construct, 
using this encoding, a symbolic {\sc Pda} that simulates the behavior of a {\sc Tpda}. 

In the classical paper by Alur and Dill on timed automata \cite{alur1994theory}, 
a \emph{region} represents a set of clock valuations with ``similar behaviors''.
The representation splits a real number into two parts: its \emph{integral value}, i.e. its value rounded down to the nearest integer,
and its \emph{fractional part}, i.e. what is left when we subtract it by its integral value. For example,
the integral value of $\pi$ is $3$, and its fractional part is $0.141592\dots$.
The main idea is that two configurations are equivalent if the following conditions hold:

\begin{itemize}
\item the integral values are identical in both valuations,
up to a constant $c_{\it max}$
\item the fractional part of any clock is either $0$ in both valuations,
or positive in both valuations 
\item the orderings of the fractional parts of all clocks are identical in both valuations
\end{itemize}

If the integral values are the same, the valuations will satisfy the same set of constraints.
If the two valuations agree on the ordering of the fractional parts, they agree on the order in which the clocks
will change integral values (and therefore in which order the constrained transitions will be enabled or disabled). 
The constant $c_{\it max}$ is the largest constant appearing syntactically in  the automaton.
All values that are above $c_{\it max}$ are indistinguishable form each other, so we can represent them symbolically
with $\omega$. In our example computation (Figure \ref{fig:tpda_example_1} and Figure \ref{fig:tpda_example_2}),
this constant is $7$.
  
\begin{figure}
\centering
\begin{tikzpicture}
\node[dummy-rsym-node,name=dummy0]
{
$
\begin{array}{c}
\tuple{\xclock_1,0}\\
\tuple{\xclock_2,3}\\
\end{array}
$
};

\node[west-rsym-node,name=r01]
at ($(dummy0.center)+(-0.5mm,0mm)$)
{$
\begin{array}{c}
\tuple{\xclock_3,1}\\
\tuple{\xclock_4,2}
\end{array}
$};

\node[west-rsym-node,name=r00] 
at ($(r01.west)+(-1mm,0mm)$)
{$
\begin{array}{c}
\tuple{\xclock_1,0}\\
\tuple{\xclock_2,2}\\
\end{array}
$};

\node[east-rsym-node,name=r02] 
at ($(dummy0.center)+(0.5mm,0mm)$)
{$
\begin{array}{c}
\tuple{\xclock_5,2}
\end{array}
$};

\node[east-rsym-node,name=r03] 
at ($(r02.east)+(1mm,0mm)$)
{$
\begin{array}{c}
\tuple{\xclock_6,3}\\
\tuple{\xclock_7,0}
\end{array}
$};

\begin{pgfonlayer}{background}
\node[stack-sym-node,name=sym1,fit=(r00) (r01) (r02) (r03)]{};
\end{pgfonlayer}

\node[fill=none, draw=none, left of=r00, xshift=-20pt] {$R_1$};
\end{tikzpicture}

\vspace{1pt}

\begin{tikzpicture}

\node[center-rsym-node,name=r12]
{$
\begin{array}{c}
\tuple{\xclock_3,1}\\
\tuple{\xclock_4,2}
\end{array}
$};

\node[west-rsym-node,name=r11] 
at ($(r12.west)+(-1mm,0mm)$)
{$
\begin{array}{c}
\tuple{\xclock_1,0}\\
\tuple{\xclock_2,2}
\end{array}
$};

\node[empty-rsym-node,name=r10] 
at ($(r11.west)+(-1mm,0mm)$)
{$
\begin{array}{c}
\tuple{\first,0}
\end{array}
$};

\node[east-rsym-node,name=r13] 
at ($(r12.east)+(1mm,0mm)$)
{$
\begin{array}{c}
\tuple{\xclock_5,2}
\end{array}
$};

\node[east-rsym-node,name=r14] 
at ($(r13.east)+(1mm,0mm)$)
{$
\begin{array}{c}
\tuple{\xclock_6,3}\\
\tuple{\xclock_7,0}
\end{array}
$};

\begin{pgfonlayer}{background}
\node[stack-sym-node,name=sym0,fit=(r10) (r11) (r12) (r13) (r14)]{};
\end{pgfonlayer}

\node[fill=none, draw=none, left of=r10, xshift=-20pt] {$R_2$};
\end{tikzpicture}

\vspace{1pt}

\begin{tikzpicture}
\node[dummy-rsym-node,name=dummy0]
{
$
\begin{array}{c}
\tuple{\xclock_1,4}\\
\tuple{\xclock_2,1}\\
\end{array}
$
};

\node[west-rsym-node,name=r01]
at ($(dummy0.center)+(-0.5mm,0mm)$)
{$
\begin{array}{c}
\tuple{\xclock_1,0}\\
\tuple{\xclock_2,2}
\end{array}
$};

\node[west-rsym-node,name=r00] 
at ($(r01.west)+(-1mm,0mm)$)
{$
\begin{array}{c}
\tuple{\xclock_6,4}\\
\tuple{\xclock_7,1}\\
\end{array}
$};

\node[east-rsym-node,name=r02] 
at ($(dummy0.center)+(0.5mm,0mm)$)
{$
\begin{array}{c}
\tuple{\xclock_3,1}\\
\tuple{\xclock_4,2}
\end{array}
$};

\node[east-rsym-node,name=r03] 
at ($(r02.east)+(1mm,0mm)$)
{$
\begin{array}{c}
\tuple{\xclock_5,2}
\end{array}
$};

\begin{pgfonlayer}{background}
\node[stack-sym-node,name=sym1,fit=(r00) (r01) (r02) (r03)]{};
\end{pgfonlayer}
\node[fill=none, draw=none, left of=r00, xshift=-20pt] {$R_3$};
\end{tikzpicture}

\vspace{1pt}

\begin{tikzpicture}

\node[center-rsym-node,name=r12]
{$
\begin{array}{c}
\tuple{\xclock_1,0}\\
\tuple{\xclock_2,2}\\
\end{array}
$};

\node[west-rsym-node,name=r11] 
at ($(r12.west)+(-1mm,0mm)$)
{$
\begin{array}{c}
\tuple{\xclock_3,4}
\end{array}
$};

\node[west-rsym-node,name=r10] 
at ($(r11.west)+(-1mm,0mm)$)
{$
\begin{array}{c}
\tuple{\xclock_6,4}\\
\tuple{\xclock_7,3}
\end{array}
$};

\node[east-rsym-node,name=r13] 
at ($(r12.east)+(1mm,0mm)$)
{$
\begin{array}{c}
\tuple{\xclock_4,2}
\end{array}
$};

\node[east-rsym-node,name=r14] 
at ($(r13.east)+(1mm,0mm)$)
{$
\begin{array}{c}
\tuple{\xclock_5,2}
\end{array}
$};

\begin{pgfonlayer}{background}
\node[stack-sym-node,name=sym0,fit=(r10) (r11) (r12) (r13) (r14)]{};
\end{pgfonlayer}
\node[fill=none, draw=none, left of=r10, xshift=-20pt] {$R_4$};
\end{tikzpicture}

\vspace{1pt}

\begin{tikzpicture}
\node[dummy-rsym-node,name=dummy0]
{
$
\begin{array}{c}
\tuple{\xclock_1,4}\\
\tuple{\xclock_2,1}\\
\end{array}
$
};

\node[west-rsym-node,name=r01]
at ($(dummy0.center)+(-0.5mm,0mm)$)
{$
\begin{array}{c}
\tuple{\xclock_1,0}\\
\tuple{\xclock_2,2}
\end{array}
$};

\node[west-rsym-node,name=r00] 
at ($(r01.west)+(-1mm,0mm)$)
{$
\begin{array}{c}
\tuple{\xclock_6,4}\\
\tuple{\xclock_7,3}\\
\end{array}
$};

\node[east-rsym-node,name=r02] 
at ($(dummy0.center)+(0.5mm,0mm)$)
{$
\begin{array}{c}
\tuple{\xclock_4,2}
\end{array}
$};

\node[east-rsym-node,name=r03] 
at ($(r02.east)+(1mm,0mm)$)
{$
\begin{array}{c}
\tuple{\xclock_3,3}\\
\tuple{\xclock_5,2}
\end{array}
$};

\begin{pgfonlayer}{background}
\node[stack-sym-node,name=sym1,fit=(r00) (r01) (r02) (r03)]{};
\end{pgfonlayer}
\node[fill=none, draw=none, left of=r00, xshift=-20pt] {$R_5$};
\end{tikzpicture}
\caption{Example regions}
\label{fig:regions_example}
\end{figure}
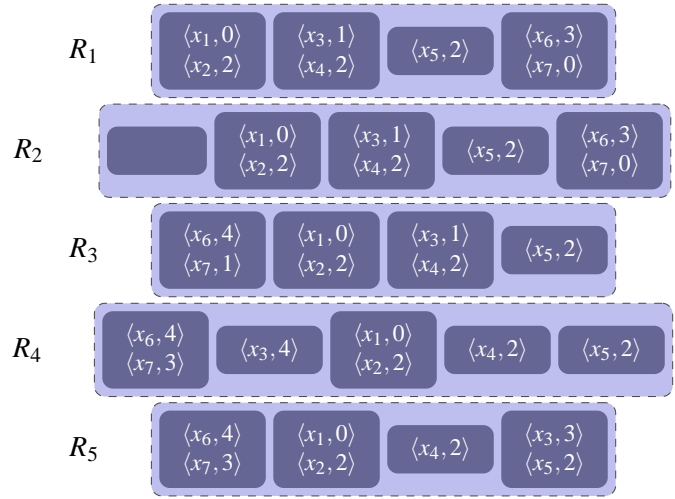

We will use a representation of regions inspired by \cite{abdulla1998verifying, abdulla2003model}, that suites our purposes.
In our representation, regions are sequences of sets. 
Each set contains one or more clocks together with their integral values. Their positioning in the sequence encodes
the ordering of the fractional parts. If two clocks are in the same set, their fractional parts are equal.
The first set contains all clocks with fractional part 0, and, for technical reasons, is the only set which may be empty.
For example, the region $R_1$ in Figure \ref{fig:regions_example} represents clock valuations in which
the values of $x_1$ and $x_2$ are exactly 0 and 2, respectively. Furthermore, the integral value of $x_3$ is 1 and the integral value of $x_4$
is 2, and so on. Finally, the clocks are ordered in the sequence by increasing fractional part.
Thus, the fractional parts of all clocks except $x_1$ and $x_2$ are strictly positive, and the fractional parts of $x_6$ and $x_7$ are 
the largest in the sequence (they are in the same set, so their fractional parts are equal).

\paragraph{Region rotations}

Given a region, we may simulate passage of time by \emph{rotating} it. 
When time passes, one of two things may happen:

\begin{itemize}
\item Some items have fractional part 0, in which case any passage of time is enough to ``push'' them out
\item No items have fractional part 0, in which case the items with the largest fractional part
reach their next integral values.
\end{itemize}

For instance, consider the region $R_2$ in Figure \ref{fig:regions_example}. The next change in the region representation
is that the values of $x_6$ and $x_7$ reach $4$ and $1$, respectively.



\section{Translation}
\label{sec:translation}

Our goal is is to reduce the reachability problem for {\sc Tpda} to the reachability problem for {\sc Pda}
by translating the given {\sc Tpda} to a {\sc Pda} which simulates it. 
We will first describe a naive approach for constructing such a {\sc Pda}. Then we show the problem with 
this approach and explain how to amend it.
At the end of this section, we show in detail how the computation in Figure \ref{fig:tpda_example_1} and
Figure \ref{fig:tpda_example_2} is simulated by the {\sc Pda}.

In the original paper on timed automata \cite{alur1994theory}, the timed automaton is simulated by a region automaton, 
i.e. a finite state automaton that encodes the regions in its states.
This abstraction relies on the fact that the set of clocks is fixed and finite. 
Since a {\sc Tpda} may in general operate on unboundedly many clocks (the stack is unbounded, and each symbol has an age), 
we cannot rely entirely on this abstraction. 

\begin{figure}[ht]
\centering
\subfloat[Run 1]{

\begin{minipage}{5cm}
\begin{tikzpicture}

\node[center-rsym-node,name=r01]
{$
\begin{array}{c}
\tuple{\asym,2}
\end{array}
$};

\node[empty-rsym-node,name=r00] 
at ($(r01.west)+(-1mm,0mm)$)
{$
\begin{array}{c}
\tuple{\asym,2}
\end{array}
$};

\node[east-rsym-node,name=r02] 
at ($(r01.east)+(1mm,0mm)$)
{$
\begin{array}{c}
\tuple{\xclock,1}
\end{array}
$};

\begin{pgfonlayer}{background}
\node[stack-sym-node,name=sym0,fit=(r00) (r01) (r02)]{};
\end{pgfonlayer}


\begin{pgfonlayer}{background}
\node[stack-node,name=stack0,fit=(sym0)]{};
\end{pgfonlayer}

\begin{pgfonlayer}{background}
\node[stack-sym-node,name=sym0,fit=(r00) (r01) (r02)]{};
\end{pgfonlayer}

\end{tikzpicture}

\vspace{2mm}

\begin{tikzpicture}


\node[center-rsym-node,name=r01]
{$
\begin{array}{c}
\tuple{\asym,2}
\end{array}
$};

\node[empty-rsym-node,name=r00] 
at ($(r01.west)+(-1mm,0mm)$)
{$
\begin{array}{c}
\tuple{\first,0}
\end{array}
$};

\node[east-rsym-node,name=r02] 
at ($(r01.east)+(1mm,0mm)$)
{$
\begin{array}{c}
\tuple{\xclock,1}
\end{array}
$};

\begin{pgfonlayer}{background}
\node[stack-sym-node,name=sym0,fit=(r00) (r01) (r02)]{};
\end{pgfonlayer}

\begin{pgfonlayer}{background}
\node[stack-node,name=stack0,fit=(sym0)]{};
\end{pgfonlayer}

\begin{pgfonlayer}{background}
\node[stack-sym-node,name=sym0,fit=(r00) (r01) (r02)]{};
\end{pgfonlayer}


\node[center-rsym-node,name=r11]
at ($(r01.north)+(0mm,8mm)$)
{$
\begin{array}{c}
\tuple{\bsym,0}
\end{array}
$};

\node[empty-rsym-node,name=r10] 
at ($(r11.west)+(-1mm,0mm)$)
{$
\begin{array}{c}
\tuple{\first,0}
\end{array}
$};

\node[east-rsym-node,name=r12] 
at ($(r11.east)+(1mm,0mm)$)
{$
\begin{array}{c}
\tuple{\xclock,1}
\end{array}
$};

\begin{pgfonlayer}{background}
\node[stack-sym-node,name=sym0,fit=(r00) (r01) (r02)]{};
\end{pgfonlayer}

\begin{pgfonlayer}{background}
\node[stack-sym-node,name=sym1,fit=(r10) (r11) (r12)]{};
\end{pgfonlayer}

\begin{pgfonlayer}{background}
\node[stack-node,name=stack1,fit=(sym0) (sym1)]{};
\end{pgfonlayer}

\begin{pgfonlayer}{background}
\node[stack-sym-node,name=sym0,fit=(r00) (r01) (r02)]{};
\end{pgfonlayer}

\begin{pgfonlayer}{background}
\node[stack-sym-node,name=sym1,fit=(r10) (r11) (r12) ]{};
\end{pgfonlayer}

\draw[tran-edge] ($(stack1.north)+(0,10mm)$) -- node[label-node,right=2pt] {$\pusht{b}{[0:1]}$} ($(stack1.north)+(0,2mm)$);

\end{tikzpicture}

\vspace{2mm}

\begin{tikzpicture}


\node[center-rsym-node,name=r01]
{$
\begin{array}{c}
\tuple{\asym,2}
\end{array}
$};

\node[empty-rsym-node,name=r00] 
at ($(r01.west)+(-1mm,0mm)$)
{$
\begin{array}{c}
\tuple{\first,0}
\end{array}
$};

\node[east-rsym-node,name=r02] 
at ($(r01.east)+(1mm,0mm)$)
{$
\begin{array}{c}
\tuple{\xclock,1}
\end{array}
$};

\begin{pgfonlayer}{background}
\node[stack-sym-node,name=sym0,fit=(r00) (r01) (r02)]{};
\end{pgfonlayer}

\begin{pgfonlayer}{background}
\node[stack-node,name=stack0,fit=(sym0)]{};
\end{pgfonlayer}

\begin{pgfonlayer}{background}
\node[stack-sym-node,name=sym0,fit=(r00) (r01) (r02)]{};
\end{pgfonlayer}


\node[dummy-rsym-node,name=dummy1]
at ($(r01.north)+(0mm,8mm)$)
{$
\begin{array}{c}
\tuple{\bsym,0}
\end{array}
$};

\node[empty-rsym-node,name=r10] 
at ($(dummy1.center)+(-0.5mm,0mm)$)
{$
\begin{array}{c}
\tuple{\first,0}
\end{array}
$};

\node[east-rsym-node,name=r12] 
at ($(dummy1.center)+(0.5mm,0mm)$)
{$
\begin{array}{c}
\tuple{\bsym,0}\\
\tuple{\xclock,1}
\end{array}
$};

\begin{pgfonlayer}{background}
\node[stack-sym-node,name=sym0,fit=(r00) (r01) (r02)]{};
\end{pgfonlayer}

\begin{pgfonlayer}{background}
\node[stack-sym-node,name=sym1,fit=(r10) (r11) (r12)]{};
\end{pgfonlayer}

\begin{pgfonlayer}{background}
\node[stack-node,name=stack1,fit=(sym0) (sym1)]{};
\end{pgfonlayer}

\begin{pgfonlayer}{background}
\node[stack-sym-node,name=sym0,fit=(r00) (r01) (r02)]{};
\end{pgfonlayer}

\begin{pgfonlayer}{background}
\node[stack-sym-node,name=sym1,fit=(r10) (r11) (r12) ]{};
\end{pgfonlayer}

\draw[tran-edge] ($(stack1.north)+(0,10mm)$) -- node[label-node,right=2pt] {$\reset{x}{[1:2]}$} ($(stack1.north)+(0,2mm)$);

\end{tikzpicture}
\end{minipage}
}
\hspace{2cm}
\subfloat[Run 2]{
\begin{minipage}{5cm}

\begin{tikzpicture}

\node[center-rsym-node,name=r01]
{$
\begin{array}{c}
\tuple{\asym,2}
\end{array}
$};

\node[empty-rsym-node,name=r00] 
at ($(r01.west)+(-1mm,0mm)$)
{$
\begin{array}{c}
\tuple{\asym,2}
\end{array}
$};

\node[east-rsym-node,name=r02] 
at ($(r01.east)+(1mm,0mm)$)
{$
\begin{array}{c}
\tuple{\xclock,1}
\end{array}
$};

\begin{pgfonlayer}{background}
\node[stack-sym-node,name=sym0,fit=(r00) (r01) (r02)]{};
\end{pgfonlayer}


\begin{pgfonlayer}{background}
\node[stack-node,name=stack0,fit=(sym0)]{};
\end{pgfonlayer}

\begin{pgfonlayer}{background}
\node[stack-sym-node,name=sym0,fit=(r00) (r01) (r02)]{};
\end{pgfonlayer}

\end{tikzpicture}

\vspace{2mm}

\begin{tikzpicture}


\node[center-rsym-node,name=r01]
{$
\begin{array}{c}
\tuple{\asym,2}
\end{array}
$};

\node[empty-rsym-node,name=r00] 
at ($(r01.west)+(-1mm,0mm)$)
{$
\begin{array}{c}
\tuple{\first,0}
\end{array}
$};

\node[east-rsym-node,name=r02] 
at ($(r01.east)+(1mm,0mm)$)
{$
\begin{array}{c}
\tuple{\xclock,1}
\end{array}
$};

\begin{pgfonlayer}{background}
\node[stack-sym-node,name=sym0,fit=(r00) (r01) (r02)]{};
\end{pgfonlayer}

\begin{pgfonlayer}{background}
\node[stack-node,name=stack0,fit=(sym0)]{};
\end{pgfonlayer}

\begin{pgfonlayer}{background}
\node[stack-sym-node,name=sym0,fit=(r00) (r01) (r02)]{};
\end{pgfonlayer}


\node[dummy-rsym-node,name=dummy1]
at ($(r01.north)+(0mm,8mm)$)
{$
\begin{array}{c}
\tuple{\bsym,0}
\end{array}
$};

\node[empty-rsym-node,name=r10] 
at ($(dummy1.center)+(-0.5mm,0mm)$)
{$
\begin{array}{c}
\tuple{\first,0}
\end{array}
$};

\node[east-rsym-node,name=r12] 
at ($(dummy1.center)+(0.5mm,0mm)$)
{$
\begin{array}{c}
\tuple{\bsym,0}\\
\tuple{\xclock,1}
\end{array}
$};

\begin{pgfonlayer}{background}
\node[stack-sym-node,name=sym0,fit=(r00) (r01) (r02)]{};
\end{pgfonlayer}

\begin{pgfonlayer}{background}
\node[stack-sym-node,name=sym1,fit=(r10) (r11) (r12)]{};
\end{pgfonlayer}

\begin{pgfonlayer}{background}
\node[stack-node,name=stack1,fit=(sym0) (sym1)]{};
\end{pgfonlayer}

\begin{pgfonlayer}{background}
\node[stack-sym-node,name=sym0,fit=(r00) (r01) (r02)]{};
\end{pgfonlayer}

\begin{pgfonlayer}{background}
\node[stack-sym-node,name=sym1,fit=(r10) (r11) (r12) ]{};
\end{pgfonlayer}

\draw[tran-edge] ($(stack1.north)+(0,10mm)$) -- node[label-node,right=2pt] {$\pusht{b}{[1:2]}$} ($(stack1.north)+(0,2mm)$);

\end{tikzpicture}
\end{minipage}
} 

\caption{Example of information loss}
\label{fig:counter-example}
\end{figure}
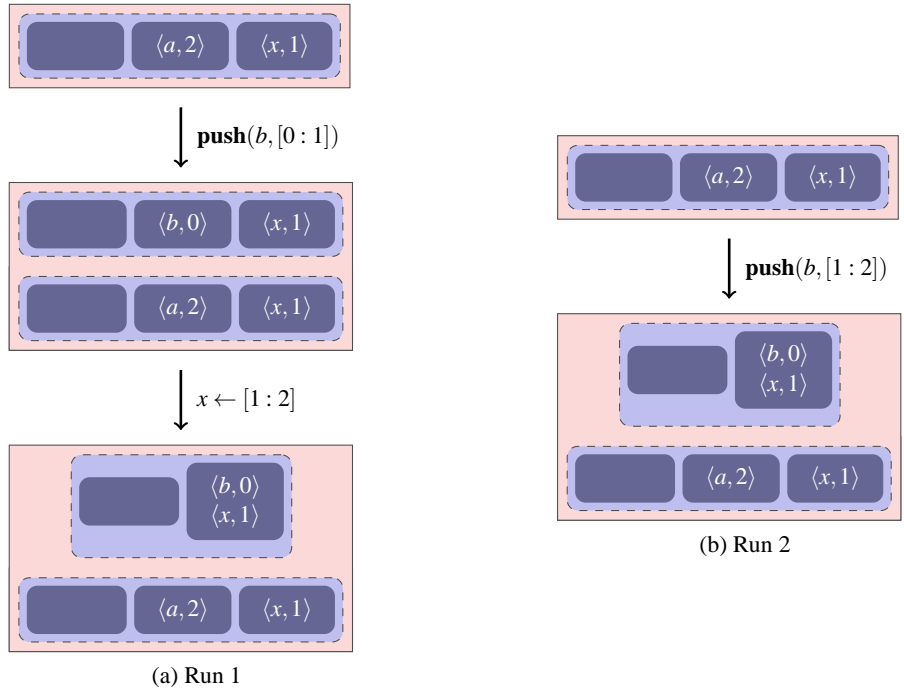

Instead, we store the regions in the stack.
Each symbol in the stack of the {\sc Tpda} is represented in the stack of the {\sc Pda} by a region that relates 
the stack symbol with all clocks. 
For example, consider Run 1 shown in Figure \ref{fig:counter-example}. 
At the beginning, the stack contains a region in which the integral values of $\asym$ and $\xclock$ are $2$  and $1$, respectively, and 
the fractional part of $\xclock$ is larger than the fractional part of $\asym$, which is in turn larger than $0$.
The {\sc Pda} then simulates the pushing of $\bsym$ with an initial age in $[0:1]$.  
This creates a new region on top of the stack which relates
$\bsym$ to $\xclock$. The region shown in the run is one of 4 possible regions. 
Next, the value of $\xclock$ is set to some value in $[1:2]$.
In our case, it happens that $\xclock$ gets the same fractional part as $\bsym$.


\begin{figure}
\centering

\begin{minipage}{5cm}
\begin{tikzpicture}


\node[center-rsym-node,name=r01]
{$
\begin{array}{c}
\tuple{\xclock,1}
\end{array}
$};

\node[empty-rsym-node,name=r00] 
at ($(r01.west)+(-1mm,0mm)$)
{$
\begin{array}{c}
\tuple{\xclock,1}
\end{array}
$};

\node[east-rsym-node,name=r02] 
at ($(r01.east)+(1mm,0mm)$)
{$
\begin{array}{c}
\tuple{\asym,2}
\end{array}
$};

\begin{pgfonlayer}{background}
\node[stack-sym-node,name=sym0,fit=(r00) (r01) (r02)]{};
\end{pgfonlayer}

\begin{pgfonlayer}{background}
\node[stack-node,name=stack1,fit=(sym0)]{};
\end{pgfonlayer}

\begin{pgfonlayer}{background}
\node[stack-sym-node,name=sym0,fit=(r00) (r01) (r02) ]{};
\end{pgfonlayer}

\end{tikzpicture}
\end{minipage}
\begin{minipage}{3.6cm}
\begin{tikzpicture}


\node[center-rsym-node,name=r01]
{$
\begin{array}{c}
\tuple{\asym,2} \\
\tuple{\xclock,1}
\end{array}
$};

\node[empty-rsym-node,name=r00] 
at ($(r01.west)+(-1mm,0mm)$)
{$
\begin{array}{c}
\tuple{\xclock,1}
\end{array}
$};

\begin{pgfonlayer}{background}
\node[stack-sym-node,name=sym0,fit=(r00) (r01)]{};
\end{pgfonlayer}

\begin{pgfonlayer}{background}
\node[stack-node,name=stack1,fit=(sym0)]{};
\end{pgfonlayer}

\begin{pgfonlayer}{background}
\node[stack-sym-node,name=sym0,fit=(r00) (r01) ]{};
\end{pgfonlayer}

\end{tikzpicture}
\end{minipage}
\begin{minipage}{5cm}
\begin{tikzpicture}


\node[center-rsym-node,name=r01]
{$
\begin{array}{c}
\tuple{\asym,2}
\end{array}
$};

\node[empty-rsym-node,name=r00] 
at ($(r01.west)+(-1mm,0mm)$)
{$
\begin{array}{c}
\tuple{\asym,2}
\end{array}
$};

\node[east-rsym-node,name=r02] 
at ($(r01.east)+(1mm,0mm)$)
{$
\begin{array}{c}
\tuple{\xclock,1}
\end{array}
$};

\begin{pgfonlayer}{background}
\node[stack-sym-node,name=sym0,fit=(r00) (r01) (r02)]{};
\end{pgfonlayer}

\begin{pgfonlayer}{background}
\node[stack-node,name=stack1,fit=(sym0)]{};
\end{pgfonlayer}

\begin{pgfonlayer}{background}
\node[stack-sym-node,name=sym0,fit=(r00) (r01) (r02) ]{};
\end{pgfonlayer}

\end{tikzpicture}
\end{minipage}

\caption{Result of popping}
\label{fig:example_pop_result}
\end{figure}
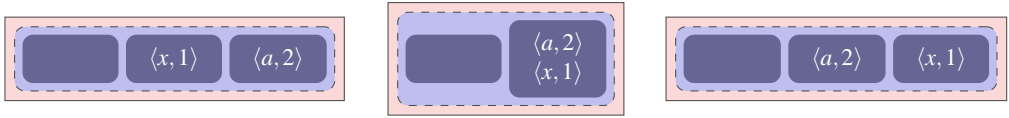

Unfortunately, it is not enough to relate each stack symbol to all clocks. 
Consider the final stack of Run 1 in Figure \ref{fig:counter-example}. 
What is the resulting stack if we now pop $\bsym$? 
It is clear that the resulting stack must contain $\asym$ and $\xclock$.
As for constraints on their values, we know from the topmost region that the fractional part of $x$ is positive.
We also know, from the region below, that the fractional part of $\asym$ is positive.
If we combine this information, we end up with one of the stacks in Figure \ref{fig:example_pop_result}.

To see the problem, consider Run 2 in \ref{fig:counter-example}. This run ends up with the same stack.
However, the fractional part of $\xclock$ in this run can not be equal to the fractional part of $\asym$, since
the value of $\xclock$ has not been reset. This rules out the stack in the middle in Figure \ref{fig:example_pop_result}.
Therefore, we need to relate the fractional parts of $\asym$ and $\bsym$.
A tempting solution is to simply record the value of $\asym$ in the region representing $\bsym$.
However, since a {\sc Pda} needs to have a finite stack alphabet, we can only record the values of finitely many 
previous stack symbols. 
At the same time, it is easy to construct counter-examples (similar to the one above) in which we need to keep
the relationship between stack symbols that lie arbitrarily far apart in the stack.
In \cite{abdulla2012dense}, we show that we can in fact enrich the regions in a finite way in order
to construct a {\sc Pda} which simulates a {\sc Tpda}. We will now explain the main points of this construction.


First, let us define the notion of items. An \emph{item} is either
a \emph{plain item} or a \emph{shadow item}.
A plain item  represents the value of a clock or the age of a stack symbol.
We add a special reference clock $\first$, which is always 0 except when simulating a pop transition. 
In other words, this reference clock is not changed when we simulate timed transitions.
Thus, the set of plain items consists of $X \cup \Gamma \cup \mset{\first}$.
On the other hand, shadow items record the values of the corresponding plain items in the region below. 
For each clock $\xclock$ and stack symbol $\asym$, the set of shadow items contains the symbols $\shxclock$ and $\shasym$.
Additionally, this set includes a shadow copy $\shfirst$ of the reference clock.
The shadow items are used to remember the amount of time that elapses while the plain items they represent are not on the 
top of the stack. A region is then represented by a sequence of sets of items.

\begin{figure}
\centering

\begin{tikzpicture}
\node[dummy-rsym-node,name=dummy0]
{
$
\begin{array}{c}
\tuple{\xclock_1,4}\\
\tuple{\xclock_2,1}\\
\end{array}
$
};

\node[west-rsym-node,name=r01]
at ($(dummy0.center)+(-0.5mm,0mm)$)
{$
\begin{array}{c}
\tuple{\shasym,1}\\
\end{array}
$};

\node[west-rsym-node,name=r00] 
at ($(r01.west)+(-1mm,0mm)$)
{$
\begin{array}{c}
\tuple{\first,0}\\
\tuple{\shfirst,0}\\
\end{array}
$};

\node[east-rsym-node,name=r02] 
at ($(dummy0.center)+(0.5mm,0mm)$)
{$
\begin{array}{c}
\tuple{\xclock_1,4} \\
\tuple{\shxclock_1,4}
\end{array}
$};

\node[east-rsym-node,name=r03] 
at ($(r02.east)+(1mm,0mm)$)
{$
\begin{array}{c}
\tuple{\xclock_1,3} \\
\tuple{\shxclock_1,3}
\end{array}
$};

\node[east-rsym-node,name=r04] 
at ($(r03.east)+(1mm,0mm)$)
{$
\begin{array}{c}
\tuple{\csym,0} \\
\end{array}
$};

\begin{pgfonlayer}{background}
\node[stack-sym-node,name=sym1,fit=(r00) (r01) (r02) (r04)]{};
\end{pgfonlayer}
\node[fill=none, draw=none, left of=r00, xshift=-20pt] {$R_2$};
\end{tikzpicture}

\vspace{1mm}

\begin{tikzpicture}
\node[dummy-rsym-node,name=dummy0]
{
$
\begin{array}{c}
\tuple{\xclock_1,4}\\
\tuple{\xclock_2,1}\\
\end{array}
$
};

\node[west-rsym-node,name=r01]
at ($(dummy0.center)+(-0.5mm,0mm)$)
{$
\begin{array}{c}
\tuple{\asym,1}\\
\tuple{\shfirst,0}\\
\end{array}
$};

\node[west-rsym-node,name=r00] 
at ($(r01.west)+(-1mm,0mm)$)
{$
\begin{array}{c}
\tuple{\first,0}\\
\end{array}
$};

\node[east-rsym-node,name=r02] 
at ($(dummy0.center)+(0.5mm,0mm)$)
{$
\begin{array}{c}
\tuple{\xclock_1,4} \\
\tuple{\shbsym_1,2}
\end{array}
$};

\node[east-rsym-node,name=r03] 
at ($(r02.east)+(1mm,0mm)$)
{$
\begin{array}{c}
\tuple{\xclock_2,3} \\
\tuple{\shxclock_1,5}
\end{array}
$};

\node[east-rsym-node,name=r04] 
at ($(r03.east)+(1mm,0mm)$)
{$
\begin{array}{c}
\tuple{\shxclock_2,3} \\
\end{array}
$};

\begin{pgfonlayer}{background}
\node[stack-sym-node,name=sym1,fit=(r00) (r01) (r02) (r04)]{};
\end{pgfonlayer}
\node[fill=none, draw=none, left of=r00, xshift=-20pt] {$R_1$};
\end{tikzpicture}

\caption{Example regions with shadow items}
\label{fig:regions_example_complete}
\end{figure}
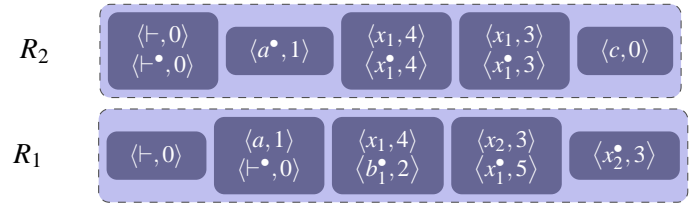

To illustrate this, let us simulate a push transition. 
Assume that the region $R_1$ in Figure \ref{fig:regions_example_complete} is the topmost region in the stack. 
The region $R_1$ records the integral values and the relationships between the clocks $x_1,x_2$, 
the topmost stack symbol $a$ and the reference clock $\first$. 
It also relates these symbols to the values of $x_1,x_2, b$ and $\first$ in the previous topmost region. 
Now, if we simulate the pushing of $c$ with initial age in $[0 : 1]$, 
one of the possible resulting regions is $R_2$. 
The region $R_2$ uses $\shxclock_1$, $\shxclock_2$ and $\shfirst$ to record the previous values of the clocks 
(initially, their values are identical to those of their plain counterparts). 
The value of the previous topmost symbol $a$ is recorded in $\shasym$. 
Finally, the region relates the new topmost stack symbol $c$ with all the previously mentioned symbols.



\paragraph{Simulation}

We will now describe how to simulate the rest of the transitions, i.e. timed transitions,
$\test{\xclock}{I}$, $\reset{\xclock}{I}$, and $\popt{\asym}{I}$.

Timed transitions are simulated by rotating the top-most region, as described in the previous section.
Note that the reference clock $\first$ is not affected by these rotations.
For example, the rotation of the topmost region between $S_{18}$ and $S_{19}$ simulates the timed transition
between $c_{18}$ and $c_{19}$ in Figure \ref{fig:tpda_example_1}.
The reference clock $\first$ stays in the first set, but all other items are rotated in a way
which is consistent with the passage of $1.75$ time units.

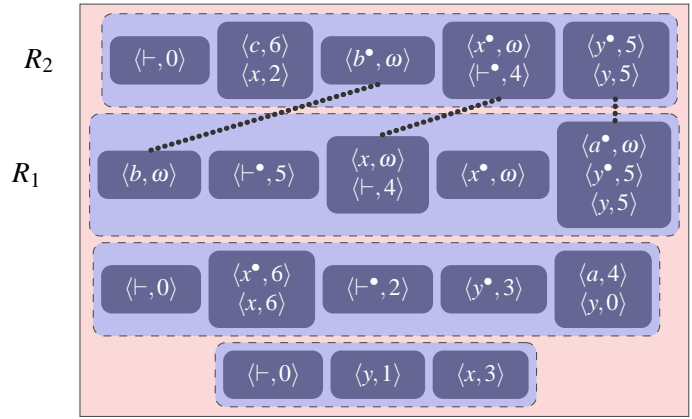
\begin{figure}
\centering
\begin{tikzpicture}


\node[center-rsym-node,name=r01]
{$
\begin{array}{c}
\tuple{\yclock,1}
\end{array}
$};

\node[west-rsym-node,name=r00] 
at ($(r01.west)+(-1mm,0mm)$)
{$
\begin{array}{c}
\tuple{\first,0}
\end{array}
$};

\node[east-rsym-node,name=r02] 
at ($(r01.east)+(1mm,0mm)$)
{$
\begin{array}{c}
\tuple{\xclock,3}
\end{array}
$};


\node[dummy-rsym-node,name=dummy1]
at ($(r01.north)+(0mm,3mm)$)
{
$
\begin{array}{c}
\tuple{\shxclock,6}\\
\tuple{\xclock,6}
\end{array}
$
};

\node[center-rsym-node,name=r12]
at (dummy1.center)
{
$
\begin{array}{c}
\tuple{\shfirst,2}
\end{array}
$
};

\node[west-rsym-node,name=r11]
at ($(r12.west)+(-1mm,0mm)$)
{$
\begin{array}{c}
\tuple{\shxclock,6}\\
\tuple{\xclock,6}
\end{array}
$};

\node[west-rsym-node,name=r10] 
at ($(r11.west)+(-1mm,0mm)$)
{$
\begin{array}{c}
\tuple{\first,0}
\end{array}
$};

\node[east-rsym-node,name=r13] 
at ($(r12.east)+(1mm,0mm)$)
{$
\begin{array}{c}
\tuple{\shyclock,3}
\end{array}
$};

\node[east-rsym-node,name=r14] 
at ($(r13.east)+(1mm,0mm)$)
{$
\begin{array}{c}
\tuple{\asym,4}\\
\tuple{\yclock,0}
\end{array}
$};


\node[dummy-rsym-node,name=dummy2]
at ($(dummy1.north)+(0mm,3mm)$)
{
$
\begin{array}{c}
\tuple{\shasym,4}\\
\tuple{\shyclock,0}\\
\tuple{\yclock,0}
\end{array}
$
};

\node[center-rsym-node,name=r22]
at ($(dummy2.center)$)
{
$
\begin{array}{c}
\tuple{\xclock,\omega}\\
\tuple{\first,4}
\end{array}
$
};

\node[west-rsym-node,name=r21]
at ($(r22.west)+(-1mm,0mm)$)
{
$
\begin{array}{c}
\tuple{\shfirst,5}
\end{array}
$
};

\node[west-rsym-node,name=r20]
at ($(r21.west)+(-1mm,0mm)$)
{$
\begin{array}{c}
\tuple{\bsym,\omega}
\end{array}
$};

\node[east-rsym-node,name=r23] 
at ($(r22.east)+(1mm,0mm)$)
{$
\begin{array}{c}
\tuple{\shxclock,\omega}
\end{array}
$};

\node[east-rsym-node,name=r24] 
at ($(r23.east)+(1mm,0mm)$)
{$
\begin{array}{c}
\tuple{\shasym,\omega}\\
\tuple{\shyclock,5}\\
\tuple{\yclock,5}
\end{array}
$};


\node[dummy-rsym-node,name=dummy3]
at ($(dummy2.north)+(0mm,3mm)$)
{
$
\begin{array}{c}
\tuple{\shasym,4}\\
\tuple{\shyclock,0}
\end{array}
$
};

\node[center-rsym-node,name=r32]
at ($(dummy3.center)$)
{
$
\begin{array}{c}
\tuple{\shbsym,\omega}
\end{array}
$
};

\node[west-rsym-node,name=r31]
at ($(r32.west)+(-1mm,0mm)$)
{
$
\begin{array}{c}
\tuple{\csym,6}\\
\tuple{\xclock,2}
\end{array}
$
};

\node[west-rsym-node,name=r30]
at ($(r31.west)+(-1mm,0mm)$)
{$
\begin{array}{c}
\tuple{\first,0}
\end{array}
$};

\node[east-rsym-node,name=r33] 
at ($(r32.east)+(1mm,0mm)$)
{$
\begin{array}{c}
\tuple{\shxclock,\omega}\\
\tuple{\shfirst,4}
\end{array}
$};

\node[east-rsym-node,name=r34] 
at ($(r33.east)+(1mm,0mm)$)
{$
\begin{array}{c}
\tuple{\shyclock,5}\\
\tuple{\yclock,5}
\end{array}
$};


\begin{pgfonlayer}{background}
\node[stack-sym-node,name=sym0,fit=(r00) (r01) (r02)]{};
\end{pgfonlayer}

\begin{pgfonlayer}{background}
\node[stack-sym-node,name=sym1,fit=(r10) (r11) (r12) (r13) (r14)]{};
\end{pgfonlayer}

\begin{pgfonlayer}{background}
\node[stack-sym-node,name=sym2,fit=(r20) (r21) (r22) (r23) (r24) ]{};
\end{pgfonlayer}

\begin{pgfonlayer}{background}
\node[stack-sym-node,name=sym3,fit=(r30) (r31) (r32) (r33) (r34)]{};
\end{pgfonlayer}

\begin{pgfonlayer}{background}
\node[stack-node,name=stack0,fit=(sym0) (sym1) (sym2) (sym3)]{};
\end{pgfonlayer}

\begin{pgfonlayer}{background}
\node[stack-sym-node,name=sym0,fit=(r00) (r01) (r02)]{};
\end{pgfonlayer}

\begin{pgfonlayer}{background}
\node[stack-sym-node,name=sym1,fit=(r10) (r11) (r12) (r13) (r14)]{};
\end{pgfonlayer}

\begin{pgfonlayer}{background}
\node[stack-sym-node,name=sym2,fit=(r20) (r21) (r22) (r23) (r24) ]{};
\end{pgfonlayer}

\begin{pgfonlayer}{background}
\node[stack-sym-node,name=sym3,fit=(r30) (r31) (r32) (r33) (r34) ]{};
\end{pgfonlayer}

\node[draw=none, fill=none, left=5mm of sym3] {$R_{2}$};
\node[draw=none, fill=none, left=5mm of sym2] {$R_{1}$};

\draw[region-edge] (r34) -- (r24);
\draw[region-edge] (r33.south) -- (r22.north);
\draw[region-edge] (r32.south) -- (r20.north);

\node[region-edge-endpoint] at (r34.south) {};
\node[region-edge-endpoint] at (r24.north) {};

\node[region-edge-endpoint] at (r33.south) {};
\node[region-edge-endpoint] at (r22.north) {};

\node[region-edge-endpoint] at (r32.south) {};
\node[region-edge-endpoint] at (r20.north) {};

\end{tikzpicture}
\caption{Simulating pop}
\label{fig:simulating_pop}
\end{figure}

The operation $\test{\xclock}{I}$ checks whether the value of $\xclock$ is in the interval $I$ or not.
For every transition rule $(s, \test{\xclock}{I}, t)$ in the {\sc Tpda} and every region that satisfies the 
condition $\xclock \in I$, we create a sequence of two transition rules which first
pops the region in question and then pushes it back. 

The reset operation $\reset{\xclock}{I}$ sets the value of clock $\xclock$ to some value in the interval $I$.
We simulate this by first popping the topmost region and then nondeterministically pushing 
a region which is identical except for the fact that $\xclock$ has been updated so that $\xclock \in I$. 
Note that there may be several regions satisfying this; the region we push is chosen nondeterministically from these.

The interesting operation is pop: the operation merges the information in two different regions.
The simulation 
is performed in two steps. 
First, the next top-most region is ``refreshed'', by repeatedly rotating it until its items are updated in a manner that 
reflects their current values. This is illustrated in Figure \ref{fig:simulating_pop}: the region $R_1$ is rotated until 
the shadow items in $R_2$ match their plain counterparts in $R_1$. In the figure, this matching is illustrated by dotted lines.
Next, we combine the regions in the following way:

\begin{itemize}
\item The plain \emph{stack symbol} is selected from the lower region ($R_1$)
\item The plain \emph{clock symbols} are selected from the upper region ($R_2$); it contains their most recent values
\item Shadow items are selected from the lower region ($R_1$)
\end{itemize}
For example, the result of combining $R_1$ and $R_2$ is the topmost region in $S_{14}$.
In this way, we simulate the passage of time only on the topmost region, but the effect
``ripples'' down the stack when popping. Thus, we only encode a finite amount of additional information in the regions,
so the stack alphabet is kept finite.

\paragraph{Results}

Given a {\sc Tpda}, we can solve the reachability problem by constructing a {\sc Pda} which simulates it,
as described in this section. The target state is reachable in the {\sc Tpda} if and only if it is reachable in the {\sc Pda}.
However, the size of the {\sc Pda} might be exponential in the size of the {\sc Tpda}. 
The following theorem states the main result in \cite{abdulla2012dense}:

\begin{theorem}
The reachability problem for {\sc Tpda} is {\sc ExpTime}-complete.
\end{theorem}

\begin{center}

\begin{figure}[h]
\centering
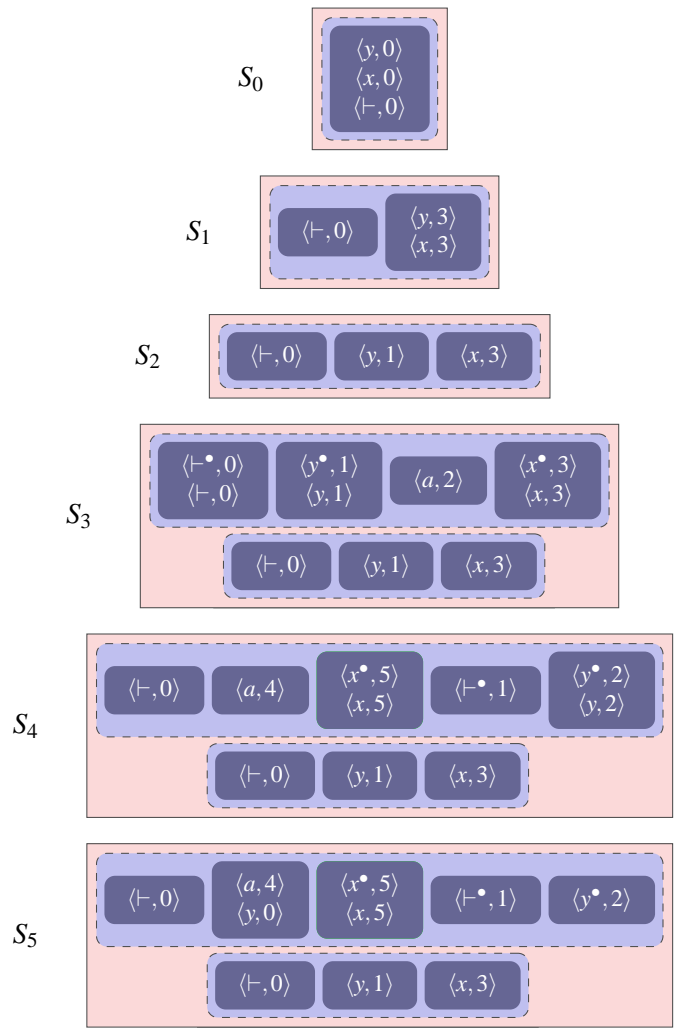
\caption{Simulation of a {\sc Tpda} computation}
\label{fig:example_simulation}


\end{center}

\bibliographystyle{eptcs}
\bibliography{biblio}

\begin{thebibliography}{1}
\providecommand{\bibitemdeclare}[2]{}
\providecommand{\surnamestart}{}
\providecommand{\surnameend}{}
\providecommand{\urlprefix}{Available at }
\providecommand{\url}[1]{\texttt{#1}}
\providecommand{\href}[2]{\texttt{#2}}
\providecommand{\urlalt}[2]{\href{#1}{#2}}
\providecommand{\doi}[1]{doi:\urlalt{http://dx.doi.org/#1}{#1}}
\providecommand{\bibinfo}[2]{#2}

\bibitemdeclare{inproceedings}{abdulla2012dense}
\bibitem{abdulla2012dense}
\bibinfo{author}{P.A. \surnamestart Abdulla\surnameend}, \bibinfo{author}{M.F.
  \surnamestart Atig\surnameend} \& \bibinfo{author}{J.~\surnamestart
  Stenman\surnameend} (\bibinfo{year}{2012}): \emph{\bibinfo{title}{Dense-timed
  pushdown automata}}.
\newblock In: {\sl \bibinfo{booktitle}{Logic in Computer Science (LICS), 2012
  27th Annual IEEE Symposium on}}, \bibinfo{organization}{IEEE},
  \doi{10.1109/LICS.2012.15}.

\bibitemdeclare{article}{abdulla2012minimal}
\bibitem{abdulla2012minimal}
\bibinfo{author}{P.A. \surnamestart Abdulla\surnameend}, \bibinfo{author}{M.F.
  \surnamestart Atig\surnameend} \& \bibinfo{author}{J.~\surnamestart
  Stenman\surnameend} (\bibinfo{year}{2012}): \emph{\bibinfo{title}{The Minimal
  Cost Reachability Problem in Priced Timed Pushdown Systems}}.
\newblock {\sl \bibinfo{journal}{Language and Automata Theory and
  Applications}}, pp. \bibinfo{pages}{58--69}, \doi{10.1007/978-3-642-28332-1}.

\bibitemdeclare{article}{abdulla1998verifying}
\bibitem{abdulla1998verifying}
\bibinfo{author}{P.A. \surnamestart Abdulla\surnameend} \&
  \bibinfo{author}{B.~\surnamestart Jonsson\surnameend} (\bibinfo{year}{1998}):
  \emph{\bibinfo{title}{Verifying networks of timed processes}}.
\newblock {\sl \bibinfo{journal}{Tools and Algorithms for the Construction and
  Analysis of Systems}}, pp. \bibinfo{pages}{298--312},
  \doi{10.1007/BFb0054179}.

\bibitemdeclare{article}{abdulla2003model}
\bibitem{abdulla2003model}
\bibinfo{author}{P.A. \surnamestart Abdulla\surnameend} \&
  \bibinfo{author}{B.~\surnamestart Jonsson\surnameend} (\bibinfo{year}{2003}):
  \emph{\bibinfo{title}{Model checking of systems with many identical timed
  processes}}.
\newblock {\sl \bibinfo{journal}{Theoretical Computer Science}}
  \bibinfo{volume}{290}(\bibinfo{number}{1}), pp. \bibinfo{pages}{241--264},
  \doi{10.1016/S0304-3975(01)00330-9}.

\bibitemdeclare{article}{alur1994theory}
\bibitem{alur1994theory}
\bibinfo{author}{R.~\surnamestart Alur\surnameend} \& \bibinfo{author}{D.L.
  \surnamestart Dill\surnameend} (\bibinfo{year}{1994}):
  \emph{\bibinfo{title}{A theory of timed automata}}.
\newblock {\sl \bibinfo{journal}{Theoretical computer science}}
  \bibinfo{volume}{126}(\bibinfo{number}{2}), pp. \bibinfo{pages}{183--235},
  \doi{10.1016/0304-3975(94)90010-8}.

\bibitemdeclare{inproceedings}{BEM97}
\bibitem{BEM97}
\bibinfo{author}{A.~\surnamestart Bouajjani\surnameend},
  \bibinfo{author}{J.~\surnamestart Esparza\surnameend} \&
  \bibinfo{author}{O.~\surnamestart Maler\surnameend} (\bibinfo{year}{1997}):
  \emph{\bibinfo{title}{Reachability Analysis of Pushdown Automata: Application
  to Model-Checking}}.
\newblock In: {\sl \bibinfo{booktitle}{CONCUR}}, \bibinfo{series}{LNCS 1243},
  \bibinfo{publisher}{Springer}, pp. \bibinfo{pages}{135--150},
  \doi{10.1007/3-540-63141-0\_10}.

\end{thebibliography}

\end{document}